\documentstyle[]{article}

\begin{document}
\begin{center} {\Large {\bf The "freely" falling two-level atom
in a running laser wave}}\\[1cm]
Karl-Peter Marzlin
\footnote{e-mail: peter.marzlin@uni-konstanz.de}
and J\"urgen Audretsch
\footnote{e-mail: juergen.audretsch@uni-konstanz.de}
\\[2mm]
Fakult\"at f\"ur Physik
der Universit\"at Konstanz\\
Postfach 5560 M 674\\
D-78434 Konstanz, Germany
\end{center}
$ $\\[3mm]
\begin{minipage}{15cm}
\begin{abstract}
The time evolution of a two-level atom which is simultaneously
exposed to the field of a running laser wave and a homogeneous
gravitational field is studied. The result of the coupled
dynamics of internal transitions and center-of-mass motion
is worked out exactly. Neglecting spontaneous emission and
performing the rotating wave approximation we derive
the complete time evolution operator in an algebraical way
by using commutation
relations. The result is discussed with respect to the
physical implications. In particular the long time and short time
behaviour is physically analyzed in detail. The breakdown of the
Magnus perturbation expansion is shown.
\end{abstract}
\end{minipage}
$ $ \\[1cm]
%%%%%%%%%%%%%%%%%%%%%%%%%%%%%%%%%%%%%%%%%%%%%%%%%%%%%%%%%%%%%%%
\section{Introduction}
Over the last ten years the manipulation of neutral atoms by
laser light was dramatically improved and has led to the new field
of atom optics (see Ref. \cite{mlynek93} and references therein).
It includes in particular atomic interferometry and laser cooling
of atoms, i.e., the preparation of a dense cloud of atoms with
a narrow momentum distribution. The width of the atomic
velocity distribution can be made as small as 1 cm/s
\cite{kastberg95}. This is of interest for atomic interferometry
because the possibility to use slow atom beams enlarges the
phase shifts caused by a broad class of external potentials
\cite{maau93}. It is obvious that for atoms moving with a velocity
of a few centimeters or meters per second for a time period
of several milliseconds or more the influence of the
earth's acceleration becomes important and cannot be neglected.
For this reason it is of interest to study the time evolution of
an atom moving in the gravitational field of the earth and the
field of a laser beam.

This time evolution is the subject of this paper. In order to
have a clear theoretical model which is exactly solvable we
restrict to the case of a two-level atom moving in a running
laser wave. Despite its relative simplicity the two-level
model is well suited for the description of certain experiments
such as Ramsey spectroscopy \cite{borde84} or atomic interferometry
\cite{riehle91,ertmer92}. But this exact solvable model
is not only accessible to experimental investigation. It may
also serve as a reference for more complicated atoms with
more than two levels which cannot be solved rigorously.
Numerical methods used in this context can be tested with the
two-level model. Another important advantage is the detailed
physical discussion which can (and will) be made using exact
solutions. The effect of every term in the Hamiltonian can easily
be identified.

We describe the dynamical behaviour of the two-level atom under the
simultaneous influence of the running laser wave and the earth's
acceleration in the Schr\"odinger picture by means of the
unitary time evolution operator
\begin{equation}
     |\psi (t) \rangle = U(t) |\psi (0) \rangle \; .
\end{equation}
which is in our case a $2\times 2$ matrix. It is methodically
essential for the following that we follow an algebraical approach
to derive $U(t)$. This approach makes only use of commutation
relations and turns out to be particularly transparent as far as
a continuous compelling physical interpretation is concerned.
Because of its independence of the initial state $|\psi_0
\rangle$ the generality of the approach is evident. Moreover,
it turns out that the resulting exact expressions for $U(t)$ are
very compact. We mention that the algebraic
method has already been used in the context of atomic
interferometry to calculate the phase shift
induced by external potentials including gravity for very general
interferometer geometries \cite{marzlin94}. The gravitational
phase shift was also calculated by Bord\'e \cite{borde92}
with a similar method.

To get a clear picture of what is going to happen it may be useful
to refer to the equivalence principle. It states that the
influence of a homogeneous gravitational field on the atom
moving in a laser wave can be simulated by constant acceleration.
This means that the following situation is physically equivalent
to our original set up: A two-level atom is at rest or moving with
constant velocity relative to an inertial system. The laboratory
with the laser attached to it moves with constant acceleration.
The consequence is that the laser wave reaches the atom with
Doppler shifted frequency. Because of the acceleration this
shift changes in time. It acts as a time dependent detuning. The
internal transitions of the atom in the limit of vanishing
acceleration and accordingly constant Doppler shift are essentially
known. For not too large detuning they are described as
Rabi oscillations. In our case, because of the time dependent
detuning, the internal behaviour of the atom in time must be
worked out anew. This is the central task of this paper. Note
that there is no further influence of gravity
on the internal dynamics of
the atom, because the fact that reference is made to an accelerated
reference frame can have no measurable consequences in this
regard.

Turning to the center-of-mass motion of the atom we know that
in the rotating wave approximation
the transitions between the two internal states are related to
the absorption or emission of one photon. The corresponding
energy and momentum transfer by recoil must show up in the
final state $|\psi(t) \rangle$. It is contained in a simple way
in the transition matrix elements $U_{12}(t)$ and $U_{21}(t)$.

Finally, there remains the motion of the atom relative to the
accelerated reference frame (or from the other point of view,
the free fall in the gravitational field). We will separate this
center-of-mass motion from the internal dynamics and the
recoil effects in factorizing $U(t)$ with $\exp (-i H_{c.m.}
t/\hbar)$ as a factor on the left. $H_{c.m.}$ is the
center-of-mass part of the Hamiltonian. Clearly this factor
drops out if one calculates the probability of finding
the atom in the excited state. This probability is not related
to the actual position and momentum of the atom. It is influenced
by gravity and acceleration only via the fact that the atom
registers the laser wave during the time $t$ with a time dependent
Doppler shift. As stated above, the additional fact that the
description refers to an accelerated reference frame does not
influence internal atomic processes.

The paper is organized as follows. In section 2 we will
introduce the model Hamiltonian and perform a unitary
transformation to get a time-independent Hamiltonian.
The exact result for the time evolution operator $U(t)$ will
be derived in section 3. Sections 4 and 5 are devoted to
a small influence of gravity and the long time limit, respectively.
A surprising and not commonly known breakdown of perturbation
theory will be demonstrated in section 6. In section 7 we will
conclude the main results and discuss possible applications.
%%%%%%%%%%%%%%%%%%%%%%%%%%%%%%%%%%%%%%%%%%%%%%%%%%%%%%%%%%%%%%%%
\section{The model}
We suppose the Hamiltonian of the two-level atom to be of the form
\begin{equation}
    H = \left (\begin{array}{cc} E_e - i \gamma_e/2 & \\ &
    E_g - i \gamma_g/2 \end{array}
    \right ) + \left (\begin{array}{cc} 1 & \\ & 1 \end{array}
    \right ) H_{c.m.} - \vec{d}\cdot \vec{E}\; \; ,
    \label{urham}\end{equation}
where $E_e$ ($E_g)$ and $\gamma_e$ ($\gamma_g$) are the energy
and the decay rate of the excited state $|\psi_e \rangle$ and the
ground state $|\psi_g \rangle$, respectively.
Although we have  inserted the decay
factors in a phenomenological way, a more complete description
of the spontaneous emission would require a density matrix
approach. For this reason the validity of our model is restricted
to times shorter than the life time of the states, but this
may be quite long for metastable states.
For convenience, we will set $\gamma_i =0$ in the calculation. This
is no loss of generality since they can be reintroduced by
replacing $E_i$ by $E_i -i \gamma_i /2$ in all expressions.
\begin{equation}
    H_{c.m.} = \frac{\vec{p}^2}{2M} - M \vec{a}\cdot\vec{x}
\end{equation}
is the center-of-mass part of the Hamiltonian with $\vec{x}$,
$\vec{p}$, and $M$ being the position and momentum operator and
the mass of the atom. $\vec{a}$ denotes the constant
gravitational acceleration acting
on the atom. The interaction with the running laser wave is
modeled by the dipole coupling
\begin{equation} \vec{d}\cdot \vec{E} = \hbar \Omega
     \cos [\omega_L t - \vec{k} \cdot \vec{x} + \varphi ]
     \left (\begin{array}{cc}  & 1\\ 1&  \end{array}
     \right )\; , \end{equation}
where $\Omega := \vec{d}\cdot \vec{E}_0/\hbar$ is Rabi's frequency.
$\vec{E}_0$ is the amplitude of the laser wave,
$\omega_L$ its frequency, $\varphi$ its phase, and $\vec{k}$ its
wave vector. $\vec{d}$ is the dipole moment of the two-level atom.

In this form the Hamiltonian depends explicitly on the time $t$.
To get rid of this dependence thus making the dynamics simple
we make a unitary transformation with the operator
\begin{equation}
    O(t)=\left (\begin{array}{cc} \exp [i(\omega_L t - \vec{k}
    \cdot \vec{x} + \varphi)/2] & \\ & \exp [-i(\omega_L t -
    \vec{k} \cdot \vec{x} + \varphi)/2] \end{array}
    \right )\label{trafo}\end{equation}
In the rotating wave approximation, i.e., after neglecting all
terms oscillating with the frequency $2 \omega_L$, we find for the
transformed Hamiltonian
$\tilde{H} = O H O^{-1} -i \hbar O\dot{O^{-1}}$ the expression
\begin{equation} \tilde{H} =
   \left (\begin{array}{cc} E_e & \\ & E_g \end{array}\right )+
   \left \{ H_{c.m.} + \frac{\hbar\delta}{4} \right \}
   \left (\begin{array}{cc} 1& \\ & 1\end{array}\right )
   - {\hbar \over 2} \{ \omega_L -\hat{D} \}
   \left (\begin{array}{cc} 1& \\ & -1\end{array}\right )
   - \frac{\hbar \Omega }{2} \left (\begin{array}{cc} & 1\\ 1 &
   \end{array}\right ) \label{htilde} \end{equation}
where we have introduced the well known recoil shift
\begin{equation} \delta := \frac{\hbar \vec{k}^2}{2M} \; .
\end{equation}

The operator
\begin{equation}
    \hat{D} := \frac{1}{M} \vec{p}\cdot \vec{k} \end{equation}
is crucial for the following calculations and their physical
implications. For an atom moving with velocity
$\vec{v}= \vec{p}/M$ it
can be written as $\vec{v}\cdot \vec{k}$. Hence this operator
represents the Doppler shift of the laser frequency in the rest
frame of the atom. We will call it the
{\em Doppler operator}.
Strictly speaking we are in this context
only allowed to argue with reference to the
transformed momentum operator
\begin{equation}
     \widetilde{\vec{p}{\bf 1}} = O \vec{p}{\bf 1}O^{-1}
     = \vec{p} {\bf 1} + {1\over 2} \hbar \vec{k} \sigma_3
     \label{pti}\end{equation}
where ${\bf 1}$ is the unit matrix in two dimensions and
$\sigma_i$ are the Pauli matrices. But with Eq.~(\ref{pti})
it is easy to see that the term containing the Doppler operator
can be written as
\begin{equation}
      \hat{D} \sigma_3 = \widetilde{\hat{D}\sigma_3} - \delta
      {\bf 1} \end{equation}
so that the interpretation is the same apart from a trivial
shift of the energy eigenvalues.

In addition to the fact that it implies a time independent
Hamiltonian, the unitary transformation (\ref{trafo}) has two
important consequences. The
momentum transfer to the center-of-mass motion related to the
absorption or emission of a photon
has been absorbed in $O(t)$. That the laser causes
internal transitions is reflected as usual by the $\Omega$-term.
The Hamiltonian (\ref{htilde}) shows very clearly that
after separation of the momentum transfer the term
containing the Doppler operator and therefore the operator
$p$ is the only one which couples
the internal degrees of freedom to the enter-of-mass motion
in a nontrivial way.
%%%%%%%%%%%%%%%%%%%%%%%%%%%%%%%%%%%%%%%%%%%%%%%%%%%%%%%%%%%%%%%%
\section{The exact time evolution operator}
The main advantage of the unitary transformation (\ref{trafo})
was to get rid of the explicit time dependence in the Hamiltonian
(after the rotating wave approximation). As a consequence the
new evolution operator $\tilde{U}(t)$ can simply be determined
by calculating
\begin{equation} O(t) U(t) O^{-1}(0)= \tilde{U}(t) =
   \exp [-i t \tilde{H}/\hbar ] =: e^{A+B}
   \label{utrafo} \end{equation}
with
\begin{eqnarray} A &:=& Q {\bf 1} \nonumber \\
     B &:=& P \sigma_3 + R \sigma_1 \end{eqnarray}
and
\begin{eqnarray}
     Q &:=& \frac{-it}{\hbar} \left \{ H_{c.m.} + {1\over 2}
     \left [ E_e + E_g + \hbar \delta/2
     \right ]\right \}\nonumber \\
     P &:=& \frac{it}{2} \{ \Delta - \hat{D} \}\nonumber \\
     R &:=& {it \over 2}  \Omega \; . \end{eqnarray}
The operators $A,B,P,Q,$ and $R$ are introduced to clarify the
mathematical structure of the calculation below.
In $P$ the quantity
\begin{equation}
      \Delta := \omega_L - \omega_{eg} \quad \mbox{with} \quad
      \omega_{eg} := \frac{E_e -E_g}{\hbar}\end{equation}
denotes the detuning of the laser frequency with respect to the
atomic transition.

To separate the free fall of the atom's center of mass we
factorize $\tilde{U}(t)$ according to
\begin{equation}
    \tilde{U}(t) = e^A W(t) \; .\label{umschreib} \end{equation}
To know the complete time development of the two-level atom
it now remains to determine the operator $W(t)$. It contains
the influence of the gravitation on the internal dynamics. It is
easy to recognize
its structure. The factorization (\ref{umschreib})
will introduce in $W(t)$ the commutator $[A,B]$ and therefore
\begin{equation} [Q,P]  =\frac{t^2}{2\hbar} [\hat{D},H_{c.m.}]
     = {i\over 2} \vec{k}\cdot \vec{a} t^2
     \label{commu}\; .\end{equation}
This is a c-number. The second equation shows that for
$\vec{k}\cdot \vec{a}\neq 0$ gravity causes a time dependent
Doppler shift. This is the central physical effect. We introduce
\begin{equation}
    \hat{D}_t := \hat{D} + \vec{k}\cdot \vec{a} t \end{equation}
(it may also be written as $\vec{k}\cdot (\vec{p}/M + \vec{a}t)$).
The relevant time scale is
\begin{equation} \tau_a := \frac{1}{\sqrt{|\vec{k}\cdot
       \vec{a}|}}\; .\end{equation}
For an
optical laser with $|\vec{k}| \approx 10^7$ m$^{-1}$ and the
earth's acceleration ($|\vec{a}| = 9.81$ m/s$^2$) $\tau_a$ is
about $10^{-4}$ seconds. Introducing
\begin{equation} \zeta := \mbox{ sgn} ( \vec{k}\cdot
     \vec{a})\end{equation}
we can rewrite Eq.~(\ref{commu}) as
\begin{equation} [Q,P] = \frac{i}{2} \zeta
       \frac{t^2}{\tau_a^2}\; .\end{equation}

To work out $W(t)$ of Eq.~(\ref{umschreib}) we employ
a method which was used by Lutzky \cite{lutzky68} in the
context of the Baker-Campbell-Hausdorff formula. We consider the
operator
\begin{equation}
   G(\lambda) := \exp [ \lambda (A+B)] =: e^{\lambda A}
   W(\lambda) \; .\label{ansatz}\end{equation}
and restrict to $\lambda=1$ at the end. Differentiation
of Eq.~(\ref{ansatz}) with respect to $\lambda$ leads to the
differential equation
\begin{equation}
   \frac{dW}{d \lambda} = \left ( e^{-\lambda A}B\; e^{\lambda A}
   \right ) W(\lambda) \label{dgl1} \end{equation}
with the initial condition $W(\lambda=0) = {\bf 1}$.
Using the identity
\begin{equation}
   e^{-\lambda A} B\; e^{\lambda A} = \sum_{n=0}^\infty
   \frac{\lambda^n}{n!} K_n \quad ,\quad K_0 := B \; , \;
   K_{n+1} := [K_n, A] \end{equation}
which holds for any two operators $A,B$ one can simplify
Eq.~(\ref{dgl1}) in our case to
\begin{equation}
   \frac{dW}{d \lambda} = \left ( B - \lambda [Q,P] \sigma_3
   \right ) W(\lambda) \label{dgl2} \; .\end{equation}
It may be written as a matrix equation,
\begin{equation}
    \left (\begin{array}{cc} dW_{11}/d \lambda & dW_{12}/d
    \lambda\\[1mm] dW_{21}/d \lambda & dW_{22}/d \lambda
    \end{array}\right ) = \left (\begin{array}{cc} P - \lambda
    [Q,P] & R \\[1mm] R & -P + \lambda [Q,P]  \end{array}\right )
    \left (\begin{array}{cc} W_{11} & W_{12} \\[1mm]
    W_{21} & W_{22} \end{array}\right )\; . \end{equation}
This is an operator-valued system of differential equations, but
it contains only commuting operators (since $[Q,P]$ is a
c-number) so that we can treat it as an ordinary differential
equation.

Inserting the equation for $dW_{11}/d \lambda$ (for
$dW_{22}/d \lambda$) into the equation for $dW_{21}/d \lambda$
(for $dW_{12}/d \lambda$) one arrives at
\begin{eqnarray}
     \frac{d^2 W_{11}}{d \lambda^2} &=& \{ R^2 -[Q,P] +
     (\lambda [Q,P] - P)^2 \} W_{11}  \nonumber \\
     \frac{d^2 W_{22}}{d \lambda^2} &=& \{ R^2 +[Q,P] +
     (\lambda [Q,P] - P)^2 \} W_{22} \; .\label{dgl3}\end{eqnarray}
After the introduction of the parameter
\begin{equation} \theta := \frac{R^2}{2[Q,P]}
     = i \frac{\Omega^2}{4\vec{k}\cdot \vec{a}}\end{equation}
and the change to the variable $y := (\lambda [Q,P] -P)
\sqrt{2/[Q,P]}$ Eq.~(\ref{dgl3}) becomes
\begin{eqnarray}
     \frac{d^2 W_{11}}{d y^2} &=& \left \{ \frac{y^2}{4} +
     \theta - {1\over 2} \right \} W_{11}(y) \label{dgl11} \\
     \frac{d^2 W_{22}}{d y^2} &=& \left \{ \frac{y^2}{4} +
     \theta + {1\over 2} \right \} W_{22}(y) \label{dgl22}
     \end{eqnarray}
with the initial conditions
$W_{11}(\lambda=0)=W_{22}(\lambda=0)=1$ and
\begin{equation}
      \frac{dW_{11}}{d y}\Bigg |_{\lambda=0} = - \frac{dW_{22}}{dy}
      \Bigg |_{\lambda=0} =
      {P\over 2} \sqrt{\frac{2}{[Q,P]}} = \frac{i}{2\sqrt{i\zeta}}
      \tau_a \hat{\Delta}_0  \end{equation}
($\hat{\Delta}_0 := \Delta - \hat{D}$).
We used the opportunity to introduce the operator of the
time dependent Doppler shifted detuning
\begin{equation} \hat{\Delta}_t := \Delta - \hat{D}_t =
     \Delta - \hat{D} - \vec{k}\cdot \vec{a} t \end{equation}
in which the effect of acceleration is already included.

The solution of Eq.~(\ref{dgl11}) and Eq.~(\ref{dgl22})is a
linear combination of
parabolic cylinder  functions. All relations between these
functions used in the following are taken from chapter 19 of
Ref.~\cite{abramo64}. The standard solutions are given by
\begin{equation}
     U(\alpha,y) = \sqrt{\pi} e^{-y^2/4} \left \{ \frac{_1\! F_1
     (\alpha /2 + 1/4,\; 1/2,\; y^2/2)}{2^{\alpha/2 +1/4}
     \Gamma(\alpha/2 +3/4)} - \frac{y\; _1\! F_1 (\alpha /2 +
     3/4,\;
     3/2,\; y^2/2)}{2^{\alpha/2 -1/4} \Gamma(\alpha/2 +1/4)}
     \right \} \label{U}\end{equation}
and
\begin{equation}
     V(\alpha ,y) = \frac{\Gamma (\alpha+1/2)}{\pi} \left \{
     \sin (\pi \alpha ) U(\alpha ,y) + U(\alpha ,-y)\right \}
     \label{V}\end{equation}
where $_1\! F_1 (\alpha, \beta, y) $ is the confluent
hypergeometric function. For $W_{11}$ we have to set
$\alpha = -1/2 +\theta$, and
for $W_{22}$ we have $\alpha = 1/2 +\theta $. The linear
coefficients
can be derived from the initial conditions. By using the
Wronskian relation
\begin{equation}
     U \frac{dV}{dy} - \frac{dU}{dy} V = \sqrt{\frac{2}{\pi}}
     \label{wronski} \end{equation}
as well as
\begin{eqnarray}
     \frac{dU}{dy}(\alpha,y) + {1\over 2}y U(\alpha,y) + (\alpha
     + {1\over 2})\, U(\alpha +1,y) &=& 0 \nonumber \\
     \frac{dV}{dy}(\alpha,y) + {1\over 2}y V(\alpha,y) -
     V(\alpha +1,y) &=& 0 \label{uvrelat} \end{eqnarray}
one deduces (and from now on we set $\lambda = 1$)
\begin{eqnarray}
     W_{11}(\vec{p}) &=& \sqrt{\frac{\pi}{2}} \left \{
     V({\scriptstyle {1\over 2}} + \theta ,{\scriptstyle]
     {-i \over \sqrt{i\zeta}}} \tau_a \hat{\Delta}_0)
     U(-{\scriptstyle {1\over 2}} +\theta ,{\scriptstyle
     {-i \over \sqrt{i\zeta}}} \tau_a \hat{\Delta}_t) +
     \theta U({\scriptstyle {1\over 2}}  +\theta ,
     {\scriptstyle {-i \over \sqrt{i\zeta}}} \tau_a
     \hat{\Delta}_0) V(-{\scriptstyle {1\over 2}} +\theta,
     {\scriptstyle {-i \over \sqrt{i\zeta}}} \tau_a
     \hat{\Delta}_t) \right\} \nonumber \\
     %%%%%%%%%%%%%%%%
     W_{12}(\vec{p}) &=& \sqrt{\frac{\pi}{2}\theta }
     \left \{ U(-{\scriptstyle {1\over 2}} +\theta ,
     {\scriptstyle {-i \over \sqrt{i\zeta}}} \tau_a
     \hat{\Delta}_0) V(-{\scriptstyle {1\over 2}} +\theta,
     {\scriptstyle {-i \over \sqrt{i\zeta}}} \tau_a
     \hat{\Delta}_t) - V(-{\scriptstyle {1\over 2}}
     +\theta  ,{\scriptstyle {-i \over \sqrt{i\zeta}}} \tau_a
     \hat{\Delta}_0) U(-{\scriptstyle {1\over 2}}+\theta
     ,{\scriptstyle {-i \over
     \sqrt{i\zeta}}} \tau_a \hat{\Delta}_t) \right \} \nonumber \\
     %%%%%%%%%%%%%%%%
     W_{21}(\vec{p}) &=& \sqrt{\frac{\pi}{2}\theta}
     \left \{ U({\scriptstyle {1\over 2}}+\theta,
     {\scriptstyle {-i \over \sqrt{i\zeta}}} \tau_a
     \hat{\Delta}_0) V({\scriptstyle {1\over 2}}+\theta,
     {\scriptstyle {-i \over \sqrt{i\zeta}}} \tau_a
     \hat{\Delta}_t) - V({\scriptstyle {1\over 2}}+\theta,
     {\scriptstyle {-i \over \sqrt{i\zeta}}} \tau_a
     \hat{\Delta}_0)U({\scriptstyle {1\over 2}}+\theta,
     {\scriptstyle {-i \over \sqrt{i\zeta}}} \tau_a
     \hat{\Delta}_t) \right \} \nonumber \\
     %%%%%%%%%%%%%%%
     W_{22}(\vec{p}) &=& \sqrt{\frac{\pi}{2}} \left \{
     \theta V(-{\scriptstyle {1\over 2}}+\theta,{\scriptstyle
     {-i \over \sqrt{i\zeta}}} \tau_a
     \hat{\Delta}_0) U({\scriptstyle {1\over 2}}+\theta,
     {\scriptstyle {-i \over \sqrt{i\zeta}}} \tau_a
     \hat{\Delta}_t) +U(-{\scriptstyle {1\over 2}}+\theta,
     {\scriptstyle {-i \over \sqrt{i\zeta}}} \tau_a
     \hat{\Delta}_0) V({\scriptstyle {1\over 2}}+\theta,
     {\scriptstyle {-i \over \sqrt{i\zeta}}} \tau_a
     \hat{\Delta}_t) \right \} \; .\label{vij} \end{eqnarray}
The time dependence is contained in $\hat{\Delta}_t$.

With Eq.~(\ref{vij}) the total operator $\tilde{U}(t)$ is given by
\begin{equation} \tilde{U}(t) = \exp \left \{ -i t \left (
     \frac{E_e + E_g}{2 \hbar} + \frac{\delta}{4}
     + \frac{1}{\hbar} H_{c.m.} \right ) \right \}
     \left (\begin{array}{cc} W_{11}(\vec{p}) & W_{12}(\vec{p})
     \\[1mm] W_{21}(\vec{p})& W_{22}(\vec{p}) \end{array} \right
     ) \; .\label{lsg}\end{equation}
It remains to cancel the initial unitary transformation $O(t)$
in Eq.~(\ref{utrafo}) to obtain the exact expression for the
time development operator:
\begin{eqnarray}
     U(t) &=& O^{-1}(t)\,  \tilde{U}(t)\,  O(0) \nonumber \\
     &=& \exp \left \{ -i t \left (
     \frac{E_e + E_g}{2 \hbar} + \frac{\delta}{2}
     + \frac{1}{\hbar} H_{c.m.} \right ) \right \}
     \left ( \begin{array}{cc} e^{-it(\omega_L - \hat{D})/2}
     e^{i \vec{k}\cdot \vec{a} t^2/4} & \\ & e^{it(\omega_L -
     \hat{D})/2} e^{-i \vec{k}\cdot \vec{a}t^2/4} \end{array}
     \right ) \times  \nonumber \\ & &
     \left (\begin{array}{cc} W_{11}(\vec{p}-\hbar\vec{k}/2)
     & W_{12}(\vec{p}-\hbar \vec{k}/2 )\; e^{i(\vec{k} \cdot
     \vec{x} - \varphi)} \\[1mm]
     W_{21}(\vec{p}+\hbar \vec{k}/2) \; e^{-i(\vec{k} \cdot
     \vec{x} - \varphi)}  &
     W_{22}(\vec{p}+\hbar \vec{k}/2 )
     \end{array} \right ) \; .\label{trlsg} \end{eqnarray}
This is the main result of the paper.
The argument $\vec{p} \pm \hbar \vec{k}/2$ in the operators
$W_{ij}$ denotes that the operator $\vec{p}$ has to be replaced by
this expression wherever it occurs in $W_{ij}$ of Eq.~(\ref{vij}).

For practical calculations it is useful to
rewrite the factor $\exp \{ -it H_{c.m.}/\hbar\}$ in
Eq.~(\ref{trlsg}) with the aid of the Baker-Campbell-Hausdorff
formula. Doing so one arrives at
\begin{equation}
     \exp \left \{\frac{-it}{\hbar} \frac{\vec{p}^2}{2M}\right \}
     \exp \left \{\frac{it}{\hbar} M \vec{a}\cdot \vec{x} \right \}
     \exp \left \{ \frac{it^2}{2\hbar} \vec{a}\cdot \vec{p}\right
     \} \exp \left \{ \frac{it^3}{3\hbar} M \vec{a}^2 \right \}
     \; . \label{cmdyn}\end{equation}
This form is especially convenient for the interpretation. The
first exponential in Eq.~(\ref{cmdyn}) describes the ordinary
kinetic energy of the atom. The second term describes the
momentum gain $M \vec{a} t$ of the accelerated atom since
the operator $\exp (i \vec{k}\cdot \vec{x})$ for any
c-number vector
$\vec{k}$ amounts in adding $\hbar \vec{k}$ to the momentum.
The third term is responsible for the displacement of the atom
by the amount $\vec{a}t^2 /2$, and the last term is a
time dependent c-number phase factor
which corrects the "error" that we have made in writing all
factors of $H_{c.m.}$ in separated exponentials.

Turning to the discussion of the result (\ref{trlsg}) we note
that the momentum transfer to the atom related to the absorption
or emission of a photon can be read off from $U_{12}$ and
$U_{21}$ (see the
exponential factor on the right). The time dependence
of $W$ goes solely back to the time dependent Doppler operator
$\hat{D}_t$ contained in $\hat{\Delta}_t$ which modifies the
detuning. $\hat{D}_t$ introduces thereby a dependence from the
center-of-mass state of the atom via the momentum operator.
Those who are familiar with the evolution
of a two-level atom in a running laser wave may miss the recoil
shift $\delta$ in $\hat{\Delta}_t$
which normally comes with the detuning \cite{borde84,maau93}.
It enters $\hat{\Delta}_t$ only after the retransformation
of $\tilde{U}$
in Eq.~(\ref{trlsg}). The replacement of $\vec{p}$ by
$\vec{p} \pm \hbar \vec{k}/2$ in $W_{ij}(\vec{p} \pm \hbar
\vec{k}/2)$ amounts in replacing
$\hat{\Delta}_t$ in  by $\hat{\Delta}_t \mp \delta$
so that the missing recoil shift is reproduced.
Rabi's frequency $\Omega$ appears in the parameter $\theta$ in the
form $\Omega \tau_a$. It grows with the intensity of the laser wave
in relation to the characteristic time $\tau_a$.

An instructive consistency check is to turn off the laser by
setting $\Omega$ and hence $\theta$ to zero. In this case the
freely falling two-level atom should be recovered. $\theta =0$
implies immediately $W_{12}= W_{21}=0$. For the evaluation
of the $W_{ii}$ it is necessary to use the identity
\cite{abramo64}
\begin{equation}
     U(-1/2,y) = e^{-y^2/4} \quad ,\quad V(1/2,y) = \sqrt{
     \frac{2}{\pi}} e^{y^2/4} \; .\label{uv2} \end{equation}
Inserting this into Eq.~(\ref{vij}) and combining the result
with Eq.~(\ref{trlsg}) leads to
\begin{equation}
     U(t) = \exp \left \{ \frac{-it}{\hbar} H_{c.m.}\right \}
     \left (\begin{array}{cc} e^{-iE_e t/\hbar} & \\ &
     e^{-iE_g t/\hbar} \end{array}\right ) \end{equation}
as was to be expected. The first term describes the free fall
of the atom, and the second term contains the internal
oscillations.
By using Eq.~(\ref{uv2}) it is also possible to derive an
expansion of Eq.~(\ref{trlsg}) for small $\Omega$. But since
the result contains various combinations of Error functions
and is not much better to interpret as the complete result
we will omit it here.

We will close this section
with a mathematical note. In the definition of
$\sqrt{i\zeta}$ the square root of a
complex factor appears. Since $\sqrt{z}$ for complex $z$ is a
multivalued function we should decide which branch has to be taken.
Fortunately, this not a big problem since the variable $y$ in
Eqs.~(\ref{U}) and (\ref{V}) (and hence $\sqrt{i\zeta}$
in Eq.~(\ref{lsg})) is squared in the argument of the
function $_1\! F_1$. There is only one linear factor of $y$ in
Eq.~(\ref{U}) which can give an additional sign.
%%%%%%%%%%%%%%%%%%%%%%%%%%%%%%%%%%%%%%%%%%%%%%%%%%%%%%%%%%%%%%%%
\section{The limit of small gravitational influence}
We now consider the limit of very large $\tau_a$, or
more precisely small $[Q,P]$. The results will thereby be
valid for all times $t$ with $t\ll \tau_a$ so that the
atom will be able to perform many Rabi oscillations.
The physical meaning of the limit is quite clear. Very
large $\tau_a$ means very small $\vec{k}\cdot \vec{a}$, i.e., the
momentum transfer from the laser beam to the atom is only
slightly altered by the acceleration because the latter is very
small or nearly perpendicular to the laser beam. The technical
details of this limit are somewhat involved and are explained in
Appendix A. The resulting expansion of $W_{ij}(\vec{p})$ up to
linear powers of $\vec{k}\cdot \vec{a}$ is
\begin{eqnarray}
 % 11-KOMPONENTE
     W_{11}(\vec{p}) &=& \cos (\hat{\omega} t/2)
     + i \frac{\hat{\Delta}_0}{
     \hat{\omega}} \sin (\hat{\omega}t/2) + \nonumber \\
     & & \frac{i}{4} \vec{k}\cdot \vec{a}t^2 \Bigg \{ -
     \frac{\hat{\Delta}_0^2}{\hat{\omega}^2}
     \left [ \cos (\hat{\omega}t/2) - \frac{\sin
     (\hat{\omega}t/2)}{\hat{\omega}t/2} \right ]
     - i\frac{\hat{\Delta}_0}{\hat{\omega}} \sin (\hat{\omega}t/2)-
     \frac{\sin (\hat{\omega}t/2)}{\hat{\omega}t/2} \Bigg \}
     \nonumber \\
% 12-KOMPONENTE
     W_{12}(\vec{p}) &=& \frac{i \Omega }{\hat{\omega}}
     \sin(\hat{\omega}t/2) - {1\over 2}
     \vec{k}\cdot \vec{a} t \frac{\Omega }{\hat{\omega}^2} \left (
     1 + \frac{it}{2} \hat{\Delta}_0 \right ) \left \{
     \cos (\hat{\omega}t/2) - \frac{\sin (\hat{\omega}t/2)}{
     \hat{\omega}t /2} \right \} \nonumber \\
% 21-KOMPONENTE
     W_{21}(\vec{p}) &=& \frac{i \Omega }{\hat{\omega}}
     \sin(\hat{\omega}t/2) + {1\over 2}
     \vec{k}\cdot \vec{a} t\frac{\Omega }{\hat{\omega}^2} \left (
     1 - \frac{it}{2} \hat{\Delta}_0 \right ) \left \{
     \cos (\hat{\omega}t/2) - \frac{\sin (\hat{\omega}t/2)}{
     \hat{\omega}t/2} \right \} \nonumber \\
% 22-KOMPONENTE
     W_{22}(\vec{p}) &=& \cos (\hat{\omega} t/2) - i \frac{
     \hat{\Delta}_0}{\hat{\omega}} \sin (\hat{\omega}t/2)+
     \nonumber \\
     & & \frac{i}{4} \vec{k}\cdot \vec{a}t^2 \Bigg \{
     \frac{\hat{\Delta}_0^2}{\hat{\omega}^2}
     \left [ \cos (\hat{\omega}t/2) - \frac{\sin
     (\hat{\omega}t/2)}{\hat{\omega}t/2} \right ]
     -i \frac{\hat{\Delta}_0}{\hat{\omega}}
     \sin (\hat{\omega}t/2) +
     \frac{\sin (\hat{\omega}t/2)}{\hat{\omega}t/2} \Bigg \}
     \label{vijentw} \end{eqnarray}
Here we have defined the frequency operator
\begin{equation} \hat{\omega} := \sqrt{ \Omega^2 +
     (-\hat{\Delta}_0)^2} \label{hom} \end{equation}
which incorporates the well known fact that the frequency of the
Rabi oscillations is altered when the laser frequency is detuned
versus the atomic transition frequency. Note that also the effect
of the Doppler shift is included in $\hat{\Delta}_0$. The
additional "--" sign before $\hat{\Delta}_0$
indicates that the negative branch of the square root has to be
taken if $\Omega$ is set equal to zero.

The analysis of Eq.~(\ref{vijentw}) is relatively simple. First, it
is easy to show that for $\vec{k}\cdot \vec{a} =0$
Eq.~(\ref{vijentw}) describes the
ordinary Rabi oscillations of a two level atom
in a running laser wave. To see this we solve Eq.~(\ref{dgl2}) with
$[Q,P]=0$ which corresponds to $\vec{k}\cdot \vec{a} =0$. The
obvious solution is
\begin{equation} W = \exp \{ \lambda B \} \; . \end{equation}
Since in the operator $B= P \sigma_3 + R \sigma_1 $ the operators
$P$ and $R$ commute one can apply the formula
\begin{equation} \exp \{-i \vec{s} \cdot \vec{\sigma}\} =
      {\bf 1} \cos (\sqrt{\vec{s}^2}) -i
     \frac{\vec{s}\cdot \vec{\sigma}}{\sqrt{\vec{s}^2}} \sin (
     \sqrt{\vec{s}^2}) \label{matrexp} \end{equation}
to reproduce just the $\vec{a}$-independent part
of Eq.~(\ref{vijentw}). This result describes in operator form
what has been obtained in
previous calculations \cite{borde84,maau93} where the time
evolution was derived (for certain wave packets) in momentum space.
It is interesting
to see the effect of the Doppler shift on the atomic
evolution. It not only alters the frequency $\hat{\omega}$ of the
Rabi floppings. A large Doppler shift also damps the transition
probability because of the $1/ \hat{\omega}$ dependence of the
matrix elements $W_{12}(\vec{p})$ and $W_{21}(\vec{p})$, and it
is (together with the detuning $\Delta$) responsible for the
imaginary part of $W_{11}$ and $W_{22}$.

Since in our model gravity alone
cannot cause transitions, the corrections
to the transition matrix elements $W_{12}$ and $W_{21}$ have
to vanish if the influence of the laser disappears, and this
is exactly what happens in Eq.~(\ref{vijentw}). We have seen
this already above. It may be
surprising why we have in this limit corrections at all,
because for the almost freely falling atom the influence of
the homogeneous gravitational field is already included in the
factor of $\exp \{-i t H_{c.m.}/\hbar \}$ in Eq.~(\ref{trlsg}).
The remaining corrections are indeed only necessary to
cancel the phase factors of $\exp \{\pm i \vec{k}\cdot
\vec{a}t^2 /4 \}$ appearing in Eq.~(\ref{trlsg}).

Like the unperturbed part the corrections for small
$\vec{k}\cdot \vec{a}$ are oscillating with the frequency
$\hat{\omega}/2$.
Furthermore, it is not difficult to see that all corrections are
bounded functions of the operator $\hat{\omega}$. This enables
us to study their time evolution. For $t=0$ they vanish. For small
$t$ the corrections to the diagonal matrix elements of $W$
grow like $t^2$, and the non-diagonal corrections grow
with $t^3$. For moderate $t$ this cubic dependence becomes
a linear envelope of an oscillating function.
The corrections are all suppressed if the Rabi frequency $\Omega$
becomes large. In this case the force caused by the laser beam
is much larger than the gravitational force so that the former
dominates the evolution. Note that the corrections
in $W_{11}$ and $W_{22}$ do not vanish for large Doppler
shifted detunings. This is a reasonable result since a large
detuning means that the laser is out of resonance and does not
affect the atomic evolution anymore. At the same time the
influence of gravity is not altered so that there must remain
some effect of gravity even when the detuning is large.

Another effect of gravity can be anticipated by examining
$W_{12}$. For vanishing corrections this is a purely
imaginary operator. Switching gravity on we see that it
develops a real part which can become large if t grows.
This is an indication that the acceleration causes
additional time dependent phase factors. We will look at
this more closely in the next section.
%%%%%%%%%%%%%%%%%%%%%%%%%%%%%%%%%%%%%%%%%%%%%%%%%%%%%%%%%%%%%%%
\section{The evolution for long times}
The results of the previous section have been obtained in
discussing the weak gravity limit of the exact result. They could
have also been worked out within an approximation scheme which
starts with the gravitation-free case and perturbs it by a
small $\vec{k}\cdot \vec{a}$. In this section
we focus on a result which is of non-perturbative character:
the behaviour of the atom at late times $t$. The formal condition
which we will use is
\begin{equation}
     |\hat{\Delta}_t|= | \hat{\Delta}_0 - \vec{k}\cdot \vec{a}
     t | \gg \frac{1}{\tau_a}\; . \label{grz} \end{equation}
In this case we can apply the asymptotic form of the functions
$U$ and $V$ in Eq.~(\ref{vij}) which depend on $\hat{\Delta}_t$.
Since $\hat{\Delta}_t$ is an operator it would be more precise
to require $| \langle \hat{\Delta}_t \rangle | \gg 1/\tau_a$ and
to consider only wavepackets for which the standard deviation of
$\hat{\Delta}_t$ remains sufficiently small.
Because any time dependence of $W_{ij}$ is enclosed in
$\hat{\Delta}_t$ it is easy to see that the condition (\ref{grz})
is equivalent to $t \gg \tau_a$ (if $\hat{\Delta}_0$ is not too
large). Physically this means that we consider the case when the
Doppler shifted detuning becomes large if it was initially
(this is $\hat{\Delta}_0$) not very large. It would be a
different physical situation if the detuning is
initially very large and becomes small due to the Doppler effect
of the accelerated atom. This will not be considered here.

The main ingredients of the limiting process in question
are explained in Appendix B. The matrix $W$ is found to be
\begin{eqnarray}
     W_{11}(\vec{p}) &\approx& \sqrt{\frac{\pi}{2}} e^{-i \zeta
     \tau_a^2 \hat{\Delta}_t^2/4}
     (-\zeta\tau_a \hat{\Delta}_t)^{-i \zeta\Omega^2 \tau_a^2/4}
     e^{\pi \Omega^2 \tau_a^2 /16} \left \{ V({\scriptstyle
     {1\over 2}}
     +\theta,{\scriptstyle {-i\over \sqrt{i\zeta}}}\tau_a
     \hat{\Delta}_0) - \frac{i \zeta U({\scriptstyle {1\over
     2}}+\theta ,{\scriptstyle {-i \over \sqrt{i\zeta}}}\tau_a
     \hat{\Delta}_0 )}{
     \Gamma (-i \zeta \Omega^2 \tau_a^2/4)} \right \} \nonumber \\
     %%%%%%%%%%%%%%%
     W_{12}(\vec{p}) &\approx& -\sqrt{\frac{i\pi \Omega^2
     }{8\vec{k}\cdot \vec{a}}}
     e^{-i \zeta \tau_a^2 \hat{\Delta}_t^2/4}
     (-\zeta\tau_a \hat{\Delta}_t)^{-i \zeta\Omega^2 \tau_a^2/4}
     e^{\pi \Omega^2 \tau_a^2 /16} \left \{ V({\scriptstyle
     {-1\over 2}}+\theta ,{\scriptstyle {-i \over
     \sqrt{i\zeta}}}\tau_a \hat{\Delta}_0) +
     \frac{4U({\scriptstyle {-1\over 2}}+\theta ,
     {\scriptstyle {-i \over \sqrt{i\zeta}}}\tau_a
     \hat{\Delta}_0)}{\Omega^2 \tau_a^2 \Gamma (-i \zeta
     \Omega^2 \tau_a^2/4)} \right \} \nonumber \\
     %%%%%%%%%%%%%%%%%
     W_{21}(\vec{p}) &\approx& \sqrt{{i \Omega^2 \over
     4\vec{k}\cdot \vec{a} }}
     e^{i \zeta \tau_a^2 \hat{\Delta}_t^2/4}
     (-\zeta\tau_a \hat{\Delta}_t)^{i \zeta\Omega^2 \tau_a^2/4}
     e^{-\pi
     \Omega^2 \tau_a^2 /16} U({\scriptstyle {1\over
     2}}+\theta,{\scriptstyle
     {-i \over \sqrt{i\zeta}}}\tau_a \hat{\Delta}_0) \nonumber \\
     %%%%%%%%%%%%%%%%
     W_{22}(\vec{p}) &\approx& e^{i \zeta \tau_a^2
     \hat{\Delta}_t^2/4}
     (-\zeta\tau_a \hat{\Delta}_t)^{i\zeta\Omega^2
     \tau_a^2/4}e^{-\pi \Omega^2 \tau_a^2 /16} U({\scriptstyle
     {-1\over 2}}+\theta,{\scriptstyle {-i \over
     \sqrt{i\zeta}}}\tau_a \hat{\Delta}_0) \; .
     \label{gt} \end{eqnarray}

The most striking feature of these operators is that
(to lowest order) only their phase varies with
$\hat{\Delta}_t$ and hence with $t$. This is not exactly true
because terms containing $\hat{\Delta}_t$ are operator valued
and can alter also the shape of a wavepacket. But for
sufficiently narrow wavepackets in momentum space the factors
containing $\hat{\Delta}_t$ in Eq.~(\ref{gt}) simply produce an
additional phase shift with logarithmic time dependence
(remember that $y^{i c} = \exp (i c \ln y)$ ). From
Eq.~(\ref{trlsg}) we then can conclude that with
growing time the matrix elements
of the total evolution operator $U(t)$ vary also only in their
phases. We will discuss this time dependence first and turn to
the amplitudes thereafter.

We interpret the result as follows:
Atoms exposed to gravity and a laser beam
may have been in resonance with the laser at some earlier
stage of their evolution. In this stage they perform
a number of Rabi oscillations. But since the momentum
dissipation due to spontaneous emission is neglected in
our model, the atoms loose during each Rabi cycle the
same amount of momentum as they gain, the net effect being
zero. During the Rabi oscillations the atoms are
accelerated by the earth's gravity so that their velocity
increases. Because of the Doppler effect the atoms are
then driven
out of resonance with the laser beam so that the Rabi
oscillations are vanishing with increasing time. The
transition between excited and ground state is frozen.
This is reflected by the fact that the absolute value
of the $W_{ij}$ is almost constant for long times.

Looking more closely to the particular matrix element
$U_{11}$ of Eq.~(\ref{trlsg})
we can write the time dependent phase factor as
\begin{equation} \exp \left \{ -i t\left (\frac{E_e+E_g}{2\hbar} +
   \frac{1}{\hbar} H_{c.m} +\frac{\delta}{2}\right )
   \right \} \exp \left \{ \frac{i t (\hat{D}-\omega_L)}{2} +
   \frac{i}{4} \vec{k}\cdot \vec{a} t^2 \right \} e^{-i\zeta
   \tau_a^2 (\hat{\Delta}_t + \delta)^2/4}
   [-\zeta\tau_a (\hat{\Delta}_t +\delta)]^{
   -i \zeta \Omega^2 \tau_a^2/4} \; .\end{equation}
Expanding the factor of $\hat{\Delta}_t$ and keeping only
time dependent terms leads us to
\begin{equation} \exp \left \{ -i t\frac{E_e}{\hbar} \right \}
     \exp \left \{ -i t\frac{1}{\hbar} H_{c.m.} \right \}
     [-\zeta\tau_a (\hat{\Delta}_t+\delta)]^{-
     i \zeta \Omega^2 \tau_a^2/4} \; . \label{zeitabh}
     \end{equation}
Here we see that each phase factor (but the last)
is linear in $t$.
The time dependence of $U_{12}$ agrees in this limit with the
one of $U_{11}$. This is reasonable because both
matrix elements correspond to atoms which are excited at
the time $t$. The result for $U_{21}$ and $U_{22}$ can be
read off from Eq.~(\ref{zeitabh}) by replacing $E_e$ by $E_g$
(these matrix elements describe atoms which are in the ground state
at time $t$) and
$[-\zeta\tau_a (\hat{\Delta}_t +\delta)]^{-i \zeta \Omega^2
\tau_a^2/4}$ by $[-\zeta\tau_a (\hat{\Delta}_t -\delta)]^{
i \zeta \Omega^2 \tau_a^2/4}$.
Each factor in Eq.~(\ref{zeitabh}) allows a proper physical
interpretation. The first exponential describes just
the internal energy of the excited state. The second is
the time evolution of a free point particle in a
homogeneous gravitational field. Hence we see that the atom
is essentially freely falling if the third term is neglected.

Nevertheless, this slowly (logarithmically) varying phase is of
physical interest since it contains both $\Omega$ and $\vec{a}$
and is therefore the only remaining
time dependent effect arising
from both gravity and laser light. Because it contains $\Omega$
its origin must be the $\Omega$-term in the Hamiltonian
(\ref{htilde}) which induces transitions between ground and
excited state. But since it is only a phase factor it does not
describe such transitions (which would be characterized by
a population
transfer between $|\psi_e \rangle$ and $|\psi_g \rangle$, i.e.,
in a time dependent change of the absolute values of $W_{ij}$).
Seen in this way the result seems to contradict our intuition,
but this contradiction can be resolved by a comparison of
the present situation with Raman transitions (see, e.g.,
Ref.~\cite{moler92}). Raman transitions are possible in a
system with two lower and one upper state ("$\Lambda$ system").
In this $\Lambda$ system one can induce direct transitions
between the two lower states without populating the upper
state by applying a laser field with a large detuning versus
the transition frequency between a lower and the upper state.
This is similar to our case: we also have a large detuning
and no population transfer to the upper state, the only difference
is that we have as a form of degeneration
only one lower state. Hence, we interpret the
third term in Eq.~(\ref{zeitabh}) as a kind of Raman transition
in a two-level system for which the large detuning prevents
the population of the upper state during a Rabi cycle. The
peculiar logarithmic time dependence of the phase is a
consequence of the time dependent detuning. We should mention that
because we have neglected spontaneous emission the argument
does not only apply to Rabi cycles of the form $|\psi_g \rangle
\rightarrow |\psi_e \rangle \rightarrow |\psi_g \rangle$ but
also to cycles of the form $|\psi_e \rangle
\rightarrow |\psi_g \rangle \rightarrow |\psi_e \rangle$.

We turn now to the discussion of the time independent part
of $W_{ij}$. The phase of this part is a rapidly oscillating
function of $\hat{\Delta}_0$. To get a feeling for
the absolute values
we consider the case of a narrow wavepacket of atoms with
an initial velocity so that $\hat{\Delta}_0$ vanishes, i.e.,
the Doppler shifted detuning is initially zero. With the aid of
the formulae in appendix B (especially Eq.~(\ref{x00})) it
is not difficult to obtain
\begin{equation} \begin{array}{lll} |W_{11}(\hat{\Delta}_0=0)|
     = & |W_{22}(\hat{\Delta}_0=0)| = &
     \sqrt{{1\over 2} (1+e^{-\pi\Omega^2
     \tau_a^2/4})} \\[2mm]
     |W_{12}(\hat{\Delta}_0=0)| = & |W_{21} (
     \hat{\Delta}_0=0)| = &
     \sqrt{{1\over 2} (1-e^{-\pi \Omega^2 \tau_a^2/4})}
     \end{array} \end{equation}
This is a surprisingly simple expression with several interesting
features. First, it only depends on the absolute value
of $\vec{k}\cdot \vec{a}$ since the variable $\zeta$ is
absent. This is reasonable since the appearance of $\zeta
\Omega^2 \tau_a^2$ in the exponent would allow imaginary absolute
values of $W_{12}$ and $W_{21}$. Second, for vanishing
laser intensity ($\Omega = 0$) we find $|W_{12}|=|W_{21}|=0$
and $|W_{11}| = |W_{22}| =1$ in accordance with the fact that
without laser beam any internal transition is impossible.
For high laser intensity
($\Omega \gg 1$) we get a complete mixing: $|W_{ij}| = 1/
\sqrt{2}$. This can be understood as a consequence of the large
frequency of the Rabi cycles. For $\Omega \rightarrow \infty$
there is
an infinite number of Rabi cycles per unit time. This is not
a well defined result. But one can replace these quick
oscillations by there average, and this procedure results of
course in an equal probability for ground and excited state
as indicated by $|W_{ij}|^2 = 1/2$.
In general the transition amplitude depends on
the variable $\Omega^2 \tau_a^2$ which can be
interpreted as representing the influence of the laser
(because of $\Omega$) acting for the time $\tau_a$
after which the atoms are essentially out of resonance.
%%%%%%%%%%%%%%%%%%%%%%%%%%%%%%%%%%%%%%%%%%%%%%%%%%%%%%%%%%%%%%%%
\section{The breakdown of the Magnus expansion}
We now have finished the physical discussion of the atomic
dynamics in a homogeneous gravitational field and a
running laser wave. This
last section of the paper is of purely theoretical interest.
We will show here that the application of the
Magnus perturbation expansion \cite{magnus54} would lead to
unphysical results for the evolution operator. This fact was
already examined in the literature (see Ref.~\cite{fernandez90}
and references therein), mostly for the harmonic oscillators
and simple two-level systems. A further analysis for the
falling two-level atom in a running laser wave has the advantage
that it is possible to use the Schr"odinger picture and that
we can compare the result of the Magnus expansion with the exact
solution obtained above.

To perform the Magnus expansion
we go back to Eq.~(\ref{dgl2}) and ask whether it is
possible to treat $[Q,P]$ as a small term and to apply a
perturbation expansion instead of solving Eq.~(\ref{dgl2})
exactly. The Magnus expansion consists of an expansion of the
exponential of an operator. Setting $W = \exp (F(\lambda))$
we may try to calculate $F$ to first order
in $[Q,P]$. This can be done by using the equation
(see, e.g., Ref.~\cite{oteo??})
\begin{equation} \frac{d}{d \lambda} e^{F(\lambda )} = \sum_{l=1}
     ^\infty \frac{(-1)^{(l+1)}}{l!} K_l e^F
     \label{allgabl} \end{equation}
with $K_1 := dF / d \lambda$ and $K_{l+1} := [K_l,F]$
which is exactly valid for any operator $F$ provided the
exponential makes sense. Eq.~(\ref{allgabl}) has the same
structure as Eq.~(\ref{dgl2}). The strategy to solve
Eq.~(\ref{dgl2}) perturbatively is therefore to find an operator
$F$ such that the sum in the r.h.s. of Eq.~(\ref{allgabl})
reproduces, to first order in $[Q,P]$, just the prefactor of $W$
in the r.h.s. of Eq.~(\ref{dgl2}). In view of this prefactor we
make the following guess for $F$:
\begin{equation}
      F = \lambda B - {1\over 2} \lambda^2 [Q,P] \sigma_3 +
     [Q,P] \sum_{n=1}^\infty \frac{(-1)^n \lambda^{n+2}}{
     (n+1)!} b_n R_n \end{equation}
where $b_n$ are numbers, $R_1 := [\sigma_3,B]$, and
$R_{n+1}:= [R_n ,B]$. Inserting this in Eq.~(\ref{allgabl})
and comparing the result (to first order in $[Q,P]$)
with Eq.~(\ref{dgl2}) leads to the condition
\begin{equation} (n+2) b_n - {1\over 2} + \sum_{l=0}^{n-2} \left (
     \begin{array}{c} n+1 \\ l+2 \end{array}\right )
     b_{n-l-1} = 0 \; ,\; n \geq 2 . \end{equation}
It is easy to check that the numbers $b_n$ are related to
the Bernoulli numbers $B_n$ by $b_n = B_{n+1}$ by using the
relation
\begin{equation} \sum_{l=0}^{n-1} \left (\begin{array}{c} n \\ l
     \end{array}\right ) B_l = 0 \; . \end{equation}
A closed expression for $F$ can be found by noting that
\begin{equation} R_{2n} = (it \hat{\omega})^{2n-2} R_2\; , \;
     R_{2n+1} = (it \hat{\omega})^{2n} R_1 \end{equation}
and $B_{2n+1}=0$ hold. Putting everything together one finds
\begin{equation} F(\lambda) = \lambda B - \frac{\lambda^2}{2} [Q,P]
     \sigma_3 + [Q,P] R_1 \frac{\lambda}{t^2 \hat{\omega}^2}
     \left \{ \frac{it \hat{\omega} \lambda}{e^{it \hat{
     \omega}\lambda}-1} -1 + \frac{i t \hat{\omega}\lambda
     }{2} \right \} + O([Q,P]^2) \; . \end{equation}
The approximate solution of our problem is obtained if
$\lambda$ is set to 1. It is obvious that this solution
becomes singular whenever the condition
\begin{equation}
    \hat{\omega} t = 2 \pi N \quad , \quad N \in {\bf N}
    \end{equation}
is fulfilled (this is possible if we neglect the decay rates
$\gamma_e$ and $\gamma_g$; if they are non-zero the solution
$F(\lambda)$ contains an unphysical resonance).

The approximation should be good for $|[Q,P]| \ll 1$, i.e.,
for $t \ll \tau_a$. But $\hat{\omega}$ is for small
detunings essentially identical to Rabi's frequency $\Omega$.
Hence the first singularity occurs at $t = 2\pi / \Omega$
which is much less than $\tau_a$ for not too small $\Omega$.
This illustrates that a perturbation approach based on the Magnus
expansion fails to describe our system properly.
This fact was explained as a consequence of a finite
convergence radius of the Magnus expansion \cite{fernandez90}.
It may be that the reason lies
in the splitting of the right hand side of Eq.~(\ref{allgabl})
into one factor $e^F$ which contains any power of $[Q,P]$,
and in the sum over $l$ which in our example is calculated only
to first order in $[Q,P]$.
%%%%%%%%%%%%%%%%%%%%%%%%%%%%%%%%%%%%%%%%%%%%%%%%%%%%%%%%%%%%%%%%
\section{Conclusions and Outlook}
In this paper we have exactly determined and
discussed the time evolution of a two-level
atom falling in a homogeneous gravitational field under the
influence of a running laser wave. The time evolution operator
has been worked out in an algebraical way. For neglected
spontaneous emission the Doppler effect was
shown to be the origin of the characteristic new physical
features. An atom which is
initially in resonance with the laser beam will first perform
Rabi oscillations. The homogeneous gravitational field
accelerates simultaneously the center-of-mass motion. During this
acceleration the Doppler shift causes the laser frequency to
drive out of resonance with the atom. Thus the Rabi oscillations
are fading away until only a slow variation of the phase
remains, no population transfer appears for large times.
This situation is similar to Raman transitions in $\Lambda$
systems.

The treatment of the model described above may be regarded as
a first step towards the inclusion of the influence of gravity
in situations of greater practical importance.
That there is (almost) no net momentum transfer of the laser to the
atom relies on the neglection of the spontaneous emission.
As long as the laser is in resonance with the atom each
Rabi cycle will be completed, no total momentum transfer occurs.
But if spontaneous emission is included Rabi oscillations can be
incomplete. The resulting momentum transfer from the laser to
the atom may be exploited to construct a gravitational atom
storage if it is adjusted so that it can cancel the acceleration
by the homogeneous gravitational field. The interruption of
the Rabi cycles can also be managed in a coherent way, e.g.,
by using a three level "$\Lambda$" scheme with two
hyperfine ground states. In this case a magnetic field
(see, e.g., Ref.~\cite{emile93})
or a microwave may be used to carry the atoms back in
their initial state. It should be mentioned that there is
already a gravitational atom trap \cite{aminoff93} which was
theoretically studied by Wallis, Dalibard, and Cohen-Tannoudji
\cite{wallis92}. The difference to our proposal is that we use
the simultaneous action of gravity and laser forces whereas in
the existing device these forces act in different time periods.
\\[3mm]
{\bf Acknowledgment}\\
It is a pleasure to thank Rainer M\"uller for discussions
on the Magnus expansion.
%%%%%%%%%%%%%%%%%%%%%%%%%%%%%%%%%%%%%%%%%%%%%%%%%%%%%%%%%%%%%%%
\begin{appendix}
\section{}
The aim of this appendix is to study the behaviour of
$U(\theta -1/2 ,y_0 + \varepsilon )$ for small
\begin{equation} \varepsilon := \sqrt{2 [Q,P]} \sim
     \frac{t}{\tau_a} \end{equation}
and to derive the corresponding limiting case of $W_{ij}$.
For brevity we have introduced the parameter
$y_0 := -i \tau_a \hat{\Delta}_0 /\sqrt{i\zeta}$.
We start with the observation that any derivative of
$U(\theta-1/2 ,y)$
with respect to $y$ can be written in the form
\begin{equation} U^{(n)}(\theta-1/2,y) = X_n(\theta -1/2 ,y)
     U(\theta-1/2 ,y) + Y_n(\theta-1/2
    ,y) U^\prime (\theta-1/2 ,y) \label{iterat} \end{equation}
where the prime denotes the derivative with respect to $y$ and the
functions $X_n$ and $Y_n$ fulfill
\begin{eqnarray}
     X_{n+1} &=& X^\prime_n + \left ( \frac{y^2}{4} + \theta
     -1/2 \right ) Y_n \nonumber \\
     Y_{n+1} &=& Y^\prime_n + X_n \label{recur} \end{eqnarray}

with $X_0 = 1$ and $Y_0 = 0$. This relation, which follows
directly from the differential equations (\ref{dgl11}) and
(\ref{dgl22}), will turn out to be
useful for the Taylor expansion of
\begin{equation} U(\theta-1/2 ,y_0 + \varepsilon ) =
     \sum_{l=0}^\infty \frac{
     \varepsilon^l }{l!} U^{(l)}(\theta-1/2 ,y_0) \end{equation}
around $y_0$. For the expansion of $X_n$ and $Y_n$ in
$\varepsilon $ it is of importance to note that
\begin{equation} y_0 = \frac{-2P}{\varepsilon } \quad ,\quad \theta
     = \frac{R^2}{\varepsilon^2 } \end{equation}
holds. Bearing this in mind one can use Eq.~(\ref{recur}) to
proof by induction that
\begin{eqnarray}
     X_{2n} &=& \varepsilon^{-2n} \left \{ (it \hat{\omega}/2)^{2n}
     \mp \frac{n}{2} \varepsilon^2 (it \hat{\omega}/2)^{2n-2} +
     O(\varepsilon^4) \right \}\nonumber \\
     Y_{2n} &=& -P
     \varepsilon^{-2n+3} n (n-1)
     \left \{ (it \hat{\omega}/2)^{2n-4} \mp \frac{(n-2)
     \varepsilon^2 }{2}
     (it \hat{\omega}/2)^{2n-6} + O(\varepsilon^4) \right \}
     \nonumber \\
      X_{2n+1} &=& -P \varepsilon^{-2n+1} n^2
     \left \{ (it \hat{\omega}/2)^{2n-2} \mp \frac{(n-1)
     \varepsilon^2 }{2}
     (it \hat{\omega}/2)^{2n-4} + O(\varepsilon^4) \right \}
     \nonumber \\
     Y_{2n+1} &=& \varepsilon^{-2n} \left \{ (it
     \hat{\omega}/2)^{2n}
     \mp \frac{n \varepsilon^2 }{2} (it \hat{\omega}/2)^{2n-2} +
     O(\varepsilon^4) \right \} \; . \label{entwick} \end{eqnarray}
is valid. Here we used the fact that
\begin{equation} {i\over 2} t \hat{\omega} = \varepsilon
     \sqrt{\theta
     + \frac{y_0^2}{4}} = \sqrt{(-P)^2 + R^2}  \end{equation}
is independent of $\varepsilon $. The operator $\hat{\omega}$ is
defined in Eq.~(\ref{hom}), and the
upper (lower) sign in Eq.~(\ref{entwick})
holds for $U(\theta-1/2 ,y_0)$ and $U(\theta+1/2,y_0)$,
respectively. Note that the same expansion holds for
$V(\theta-1/2 ,y_0)$ and $V(\theta+1/2, y_0)$ simply because
they are also solutions of
Eqs.~(\ref{dgl11}) and (\ref{dgl22}), respectively, and because
these differential equations are the only relation which were used
to derive Eq.~(\ref{recur}).

We are now ready to calculate the expansion of $W_{12}(\vec{p})$
in terms of $\varepsilon $. We have with Eq.~(\ref{vij})
\begin{eqnarray} W_{12}(\vec{p}) &=& \sqrt{\frac{\pi}{2}}
     \frac{R}{\varepsilon }\left \{
     U(\theta-1/2 ,y_0) V(\theta -1/2 ,y_0 + \varepsilon ) -
     V( \theta -1/2 ,y_0) U(\theta -1/2 ,y_0 + \varepsilon )
     \right \} \nonumber\\
     &=& \sqrt{\frac{\pi}{2}} R \sum_{n=0}^\infty \frac{
     \varepsilon^{n-1}}{n!} Y_n(\theta -1/2 ,y_0)
     \{ U(\theta -1/2 ,y_0) V^\prime(\theta -1/2
     ,y_0) - U^\prime(\theta -1/2 ,y_0) V(\theta -1/2 ,y_0) \}
     \nonumber \\ &=& R \sum_{n=0}^\infty \frac{\varepsilon^{n-1}
     }{n!} Y_n(\theta -1/2 ,y_0) \end{eqnarray}
where we have used Eq.~(\ref{iterat}) for the first step and
the Wronskian relation (\ref{wronski}) for the second step.
It is now straightforward to derive a closed expression for
$W_{12}$ by exploiting Eq.~(\ref{entwick}) and the Taylor series
of cos and sin. The result is given in Eq.~(\ref{vijentw}).

The derivation of the expansion for $W_{21}$ is similar to
the previous calculations and will not be reproduced here.
For $W_{11}$ and $W_{22}$ one further step has to be
included since the functions $U$ and $V$ occurring in the
corresponding expressions of Eq.~(\ref{vij}) contain both
the parameter $\theta -1/2 $ and $\theta +1/2$. It is therefore
necessary to make use of Eq.~(\ref{uvrelat}) and to deduce
expressions for the $W_{ii}$ where either only
$\theta -1/2 $ or only $\theta +1/2 $ occurs. After this has been
done the calculations are the same as for $W_{12}$.
Inserting all definitions one arrives at Eq.~(\ref{vijentw}).
%%%%%%%%%%%%%%%%%%%%%%%%%%%%%%%%%%%%%%%%%%%%%%%%%%%%%%%%%%%%%%%%%%
\section{}
In this appendix we will sketch the derivation of
the long time behaviour for the
operators $W_{ij}(\vec{p})$ in Eq.~(\ref{vij}).
This can be handled
by using (Eq. 13.5.1 of Ref. \cite{abramo64})
\begin{equation}
     _1\! F_1 (\alpha ,\beta ,z) \approx \frac{\Gamma (\beta)
     }{\Gamma (\beta -\alpha)} e^{\pm i \pi \alpha} z^{-\alpha}
     + \frac{\Gamma (\beta )}{\Gamma (\alpha )} e^z z^{\alpha
     -\beta } \label{1f1naeh} \end{equation}
where $z$ is a complex number with $|z| \gg 1$, and the upper
sign holds for $-\pi /2 <$ arg $z < 3\pi /2 $ whereas the
lower sign holds for $-3 \pi /2 <$ arg $z \leq -\pi /2 $.
We apply this formula to the function
$U(-1/2+\theta ,-i \tau_a \hat{\Delta}_t /\sqrt{i\zeta})$
and in the same manner to any other function in
Eq.~(\ref{vij}) which depends on $\hat{\Delta}_t$ by
inserting it into
Eqs.~(\ref{U}) and (\ref{V}). Note that the limit depends
on $\zeta $ since $\zeta =\pm 1 $ determines arg $z$.
We assume $\langle -\tau_a \hat{\Delta}_t \rangle$ to be
a large positive number and choose arg
$(i \zeta) = \zeta \pi /2 $. Hence the "+" sign in
Eq.~(\ref{1f1naeh})
holds for $\zeta =1$ and the "--" sign for $\zeta =-1$.

The rest of the calculation is straightforward but long.
It is useful to apply $z \Gamma (z) = \Gamma (z+1)$,
$\Gamma (1/2) = \sqrt{\pi}$, and
\begin{eqnarray} \Gamma (i x) \Gamma (-ix) &=& |\Gamma (ix)|^2 =
     \pi / [ x \sinh (\pi x)] \nonumber \\
      \Gamma (i x+1/2 ) \Gamma (-ix+1/2) &=& |\Gamma
     (ix+1/2)|^2 = \pi / \cosh (\pi x) \nonumber \\
     \Gamma (2z) &=& \Gamma (z) \Gamma (z+1/2) 2^{2z -1/2}
     /\sqrt{2\pi} \end{eqnarray}
(for real $x$, see chapter 6 of Ref. \cite{abramo64})
to handle the
various factors of the $\Gamma $ function arising in the
derivation. For the case of vanishing detuning the following
equations are of use:
\begin{eqnarray}
     U(\alpha ,0) &=& \frac{\sqrt{\pi}}{2^{\alpha /2 +1/4} \Gamma
     (3/4 + \alpha/2)} \nonumber \\
     V(\alpha ,0) &=& \frac{2^{\alpha /2 +1/4}\sin [ \pi (
     \alpha+1/2)/2]}{ \Gamma (3/4 -\alpha/2)} \; . \label{x00}
     \end{eqnarray}
\end{appendix}

\end{document}